\documentclass[breakmath, report, preprint]{seismica}
\usepackage[normalem]{ulem}

\title{The BRUIT-FM "open data" noise reduction challenge}
\shorttitle{BRUIT-FM open data challenge} 

\reporttype{Data Report} 

\author[1,2]{Wayne~C~Crawford
	\orcid{0000-0002-3260-1826}
	\thanks{Corresponding author: crawford@ipgp.fr}
}
\author[3]{Stephan~Ker
	\orcid{0000-0002-4393-3965}
}
\author[3]{Simon~Rebeyrol
	\orcid{0000-0002-2568-4993}
}
\author[4]{Laurent~Duval
	\orcid{0000-0002-7732-4666}
}
\affil[1]{Institut de Physique du Globe de Paris, CNRS UMR 7154, Université Paris Cité, Paris, France}
\affil[2]{Laboratoire Littoral, Science, et Sociétés, CNRS UMR 7154, La Rochelle University, La Rochelle, France}
\affil[3]{Geo‐Ocean, UMR 6538, Université de Brest, CNRS, Ifremer, Plouzané, France}
\affil[4]{IFP Energies nouvelles, Rueil-Malmaison, France}

\credit{Conceptualization}{Crawford, Ker, Rebeyrol}
\credit{Methodology}{Crawford, Ker, Rebeyrol}
\credit{Software}{Crawford}
\credit{Writing - Original draft}{Crawford}
\credit{Writing - Review \& Editing}{Ker, Duval}

\begin{document}

	
	\makeseistitle{
		\begin{summary}{Abstract}
            The BRUIT-FM Challenge asks participants to reduce "noise" --- defined as any signal that masks the signal of interest --- on an open dataset of real and synthetic broadband seafloor seismology data. 
            The dataset includes signals from earthquakes, ocean infragravity waves and seafloor currents.
            It also includes training datasets of synthetic data with reference solutions.
            Participants are asked to remove or separate these signals to obtain the best earthquake and seafloor compliance signals.  The authors will make a first assessment of the submissions, based on
             noise reduction, signal deformation and ease of use.
            Participants are invited to a community workshop, in which they will validate the
            preliminary assessment and collaborate on a community paper describing the results, identifying promising pathways and proposing new directions.  The best results will be added to a new version of the dataset, which will persist as a benchmark for future developments.
		\end{summary}
	}  
		
	\section{Introduction}

    Seafloor seismology measurements are needed to study the structure and dynamics of Earth features that are beneath or near oceans and seas  (i.e. subduction zones, mid-ocean ridges, oceanic volcanos, and coastal fault zones) and to properly cover the earth for global seismology studies.  The water layer over the measurement stations affects the seismology signal (e.g. \citep{ruan2014, blackman1995}) and creates new signals --- such as the motion under pressure from ocean waves (\citep{yamamoto1983, crawford1991}) and the tilt generated by seafloor currents (e.g. \citep{crawfordwebb2000}). To best use one of these signals, the other signals should be separated or minimized (e.g. \citep{duennebiersutton1995, webbcrawford1999, crawfordwebb2000, janiszewski2023}).  Several computational toolboxes have been created to reduce seafloor noise, primarily at frequencies below the microseism peak (< 0.03 Hz).  Most of them are based on the calculation of the transfer function with another channel that contains a stronger version of the noise signal (e.g. \cite{janiszewski2019, crawfordwebb2000, rebeyrol2024, aminian2025}), but there are also  methods inspired by source separation algorithms designed for musical signals \citep{zali2023} and re-orientation of the vertical channel \citep{harmon2022}.  Marine seismologists are thus confronted by an array of tools, without knowing which is the best for their given problem, because neither the efficiency of each method, nor the potential for them to deform the target signal, has been objectively evaluated.
    
    We have created a dataset and a protocol to evaluate the efficiency of noise reduction techniques and toolboxes.  Our goal is to identify which existing techniques work the best and how new techniques might allow us to further improve noise reduction.  
    
    The dataset consists of two "challenge" data files ("validation", in machine learning language) and a large number of "training" data files and their answer files ("calibration").  We ask participants to run their algorithms on the challenge datasets and to return answer files to our open source repository (see section \ref{section:submitting}).  For scientists who are not used to seafloor seismology datasets, we also provide sample codes for reading and writing the files, and compliance may be returned in its "raw" (digital data values) format, without taking into account the instrument response. 

    We will first collect the answers that were generated using different techniques and compare the outputs using three criteria: 1) noise reduction, 2) deformation of the desired signal, and 3) ease of use.
    The codes we use to evaluate these parameters are provided in the dataset, as explained in section \ref{evaluation_example}.  Preliminary results will be shared, using private nicknames provided by the particpants, on the BRUIT-FM website.

We will then organize a community workshop to evaluate our classification, create other validation metrics if needed, discuss the advantages and disadvantages of each method, explore the possibility of data improvement at higher and lower frequencies, and write a community paper on the results.  All codes and datasets that have positive results using the final metrics will be integrated into the open dataset, as well as the updated evaluation tests, to allow future tests of new techniques.

    \section{The Datasets}

    The datasets are available from a Zenodo repository \citep{crawford_2025_bruitfmdata}. There are two challenge datasets, one using real data (8 days) and the other
    using synthetic data (10 days).  
    There are also one main (10-day) and 100 "mass" (1-day) training datasets.
    Each training dataset includes reference answer ("calibration") files for the outputs.
      
    \subsection{Data format and sampling rate}
    The data are stored in miniSEED files with four channels named: \code{LDG} for the pressure component, \code{LH1} and \code{LH2} for the horizontal seismometer components, and \code{LHZ} for the vertical seismometer component.  The sampling rate is 1 sample per second. Provided example scripts show how to read the data and write the outputs.
       
    \subsection{Synthetic Data}

    Synthetic data allows us to generate reference answer ("calibration") files that can be used to evaluate results.
    They are constructed using a combination of 1) real seismological data from a quiet land station and 2) synthetic seafloor noise data constructed using the \code{SeafloorSynthetic} class of the \code{tiskitpy} software
    module \citep{crawford_2025_tiskitpy}.  
    
\begin{figure*}[ht!]
		\centering
		\includegraphics[width=\textwidth]{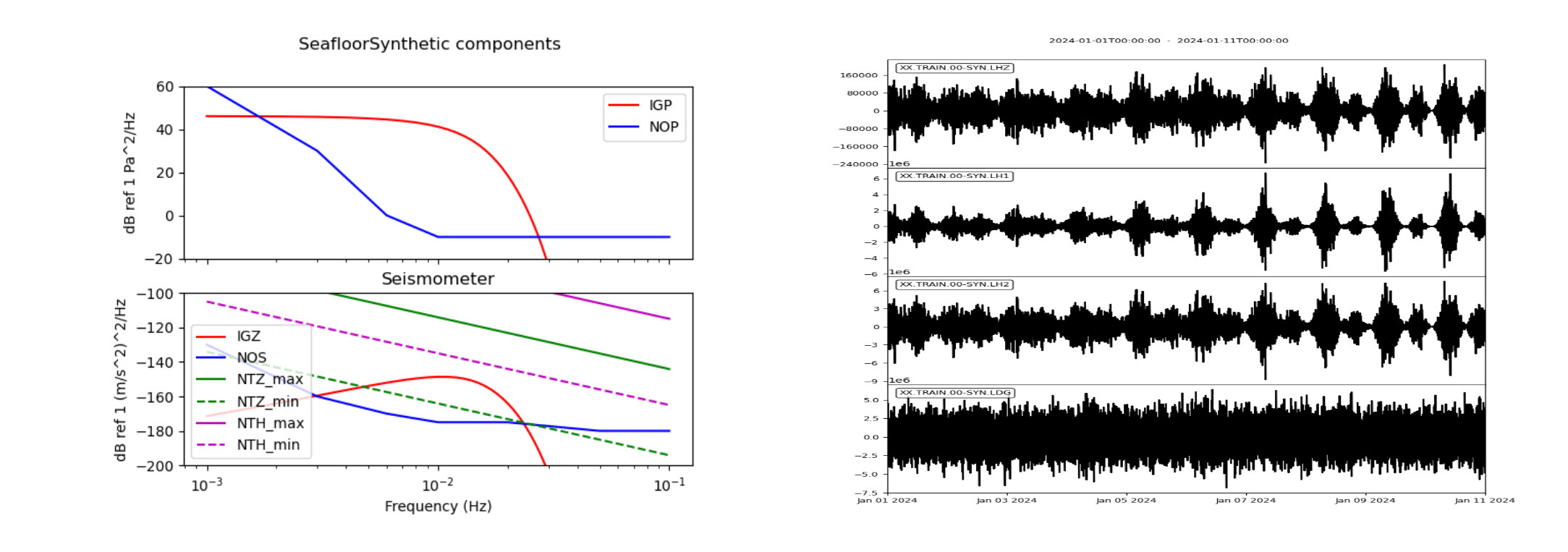} 
		\caption{An example synthetic dataset.  Left: Input power spectral density components.  (top): The pressure channel's infragravity wave signal (IGP) and sensor noise floor (NOP). (bottom): The seismometer channels' vertical motion under infragravity wave pressure (IGZ), seismometer noise floor (NOS) and maximum/minimum tilt noise on the horizontal channels (NTH\_max/min) and vertical channel (NTZ\_max/min)   Right: Channel time series. (top): vertical seismometer. (middle): Two horizontal seismometer channels, (bottom): Pressure.}
		\label{fig:noise_spect}
	\end{figure*}

    The \code{SeafloorSynthetic} class simulates seafloor noise as a combination of (Fig.~\ref{fig:noise_spect}):
     \begin{enumerate}
       \item Instrumental noise floors for the seismometer and pressure sensors;
       \item The infragravity (low frequency) wave surface height;
       \item A 1D model of earth structure beneath the seafloor station;
       \item The misalignment of the seismometer's vertical channel from gravitational vertical;
       \item The maximum seismometer tilt noise levels at the seafloor;
       \item The direction of the tilt noise with respect to the seismometer;
       \item The number of decibels below the maximum tilt noise levels for minimum tilt levels; and
       \item The water depth.
     \end{enumerate}
     It does not include the ocean wind wave surface spectrum or microseisms, which are important at higher frequencies and/or shallow (<500m) water depth. It does not model temporal variations in the infragravity wave energy, nor in the seismometer's misalignment.
     The class generates seafloor pressure and acceleration signals from the parameters, varying the tilt noise over time in rhythm with the tides.

    \subsection{Challenge Datasets}
    \subsubsection{Real Data}

    The real data are from a seafloor station with high local seismicity and a relatively small compliance signal.
     An example of the outputs obtainable using the \code{tiskitpy} module is provided with the dataset and shown in  subsection \ref{cleaningexample}.

   \subsubsection{Synthetic Data}

    The synthetic data are the sum of the waveforms from synthetically generated noise and real seismograms from an unspecified broadband land station.  The real seismograms include significant global earthquakes, but we manually reset the data start date to 2024-01-01, to hide which earthquakes these are and the real station's distance from them.

    \subsection{Training Datasets}
    These datasets are generated by adding synthetic seafloor noise to data from a quiet land station.  Reference ("calibration") answer files are provided for each dataset: \code{*_REF_clean_eq.mseed} is the original land station data, \code{*_REF_clean_compl.mseed} contains timeseries based on only the compliance component of the synthetic noise and \code{*_REF_compliance_*.csv} is the compliance values input to generate this part of the synthetic noise.
    
    \subsubsection{Main dataset}

    The main training dataset is generated by the same process as the synthetic challenge dataset, except with different noise parameters and a different record of high global seismicity recorded on a quiet land station.

   \subsubsection{Mass datasets}

    The mass training dataset consists of 100 synthetic data files and their answer ("calibration") files.
    The synthetic data are generated in the same way as for the synthetic challenge and main training waveforms, with noise parameters and the time period of the land data randomly selected. The land station data for half of the mass datasets includes a global earthquake  with a reported magnitude greater than or equal to 5.2.
       
    \subsection{Codes}
    The dataset also includes Python scripts showing how to read the data and writing the answer files and to package answers for submission.  It also includes the scripts we use to evaluate the challenge responses, with instructions on how to use them on the training data.
    
 \section{Seafloor compliance} \label{compliance}
    For the purposes of this challenge, seafloor compliance is the ratio of the vertical seafloor motion to the pressure signal, at frequencies where the two are coherent.  "Cleaned for compliance" data are data from which all noise sources that reduce the pressure-vertical coherence have been removed/minimized.  Existing methods for creating "cleaned for compliance" data are described in \citep{crawfordwebb2000, janiszewski2019}, using transfer functions, and \cite{harmon2022}, using tilt corrections.
    
    Compliance can be returned either in units of the digital data or in Pa$^{-1}$: the latter requires using instrument response and water depth information from the inventory file.  More details on seafloor compliance can be found in \cite{yamamoto1983, crawford1991, crawford1998, aminian2025}.

  \section{The noise reduction challenge}

   Participants should download the dataset from the Zenodo repository \citep{crawford_2025_bruitfmdata}.
   They should apply their noise reduction codes to the challenge datasets and create a ZIP file containing one or more of the following files:
   \begin{enumerate}
       \item Waveform data cleaned of everything except earthquakes ("cleaned for earthquakes");
       \item Waveform data cleaned of everything except seafloor compliance ("cleaned for compliance");
       \item Seafloor compliance, in units of data (COUNTS) or Pa$^{-1}$;
       \item Source code or an explanation of the algorithms used.   
   \end{enumerate}
   File names and formats are specified in section \ref{fileformats}, training datasets with reference outputs are in the \code{TRAINING/} subdirectories and our evaluation codes are provided in the \code{CODES/evaluation} directory.  The answer and manifest files should be packaged into a ZIP file and submitted to our open-source  \href{https://form.jotform.com/262153865166058}{Jotform repository}, and a simplified results file can also be submitted to our \href{https://kaggle.com/competitions/the-bruit-fm-open-data-noise-reduction-challenge}{Kaggle.com site} (requires an account).  Codes for evaluating the training results and generating the submission files are included with the dataset. Questions may be addressed to: \href{mailto:bruitfm_challenge@services.cnrs.fr}{bruitfm\_challenge@services.cnrs.fr}.
 
   No knowledge of seismological processing is needed.  The data are in a seismological standard format (miniSEED), but example code is provided (using the open source \code{obspy} \citep{beyreuther2010} software) to read and display the data.  Each directory contains an "inventory" file in StationXML format, which is only needed if the participant chooses to return the compliance in units of Pa$^{-1}$.
      
        \subsection{An example of cleaning the main training dataset} \label{cleaningexample}

\begin{figure*}[ht!]
		\centering
		\includegraphics[width=\textwidth]{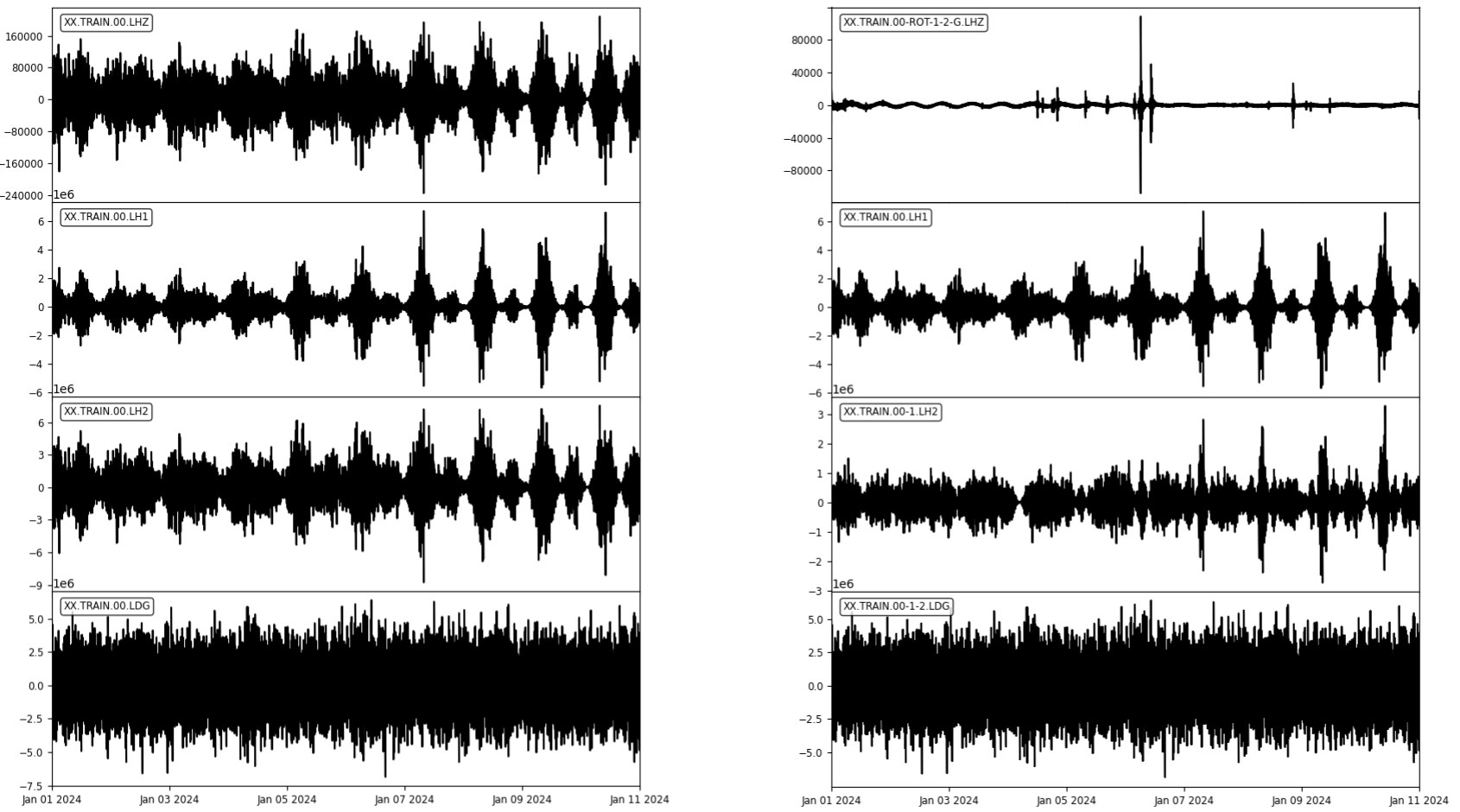} 
		\caption{Waveforms.  Left: Original main training data.  Right: "Cleaned for earthquakes" data.}
		\label{fig:code_3_waveforms}
	\end{figure*}

\begin{figure*}[ht!]
		\centering
		\includegraphics[width=8.6cm]{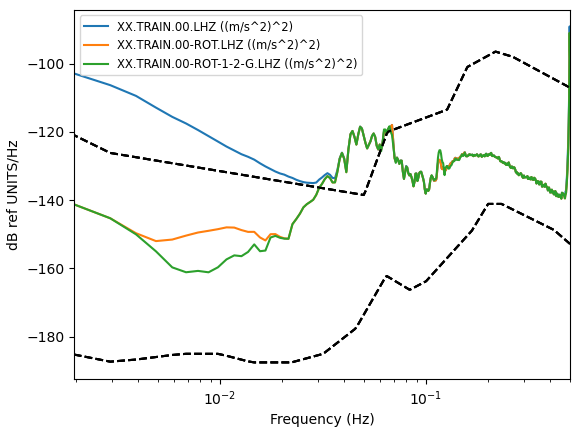} 
		\caption{Vertical channel power spectral densities for the main training data: original (blue), cleaned for compliance (orange), cleaned for earthquakes (green).}
		\label{fig:code_3_PSDs}
	\end{figure*}

    We show below the results of the repository's example code \code{CODES/step3\_read\_write\_clean\_tiskitpy.py}, which reads in the main training dataset, cleans it using \code{tiskitpy}, and outputs the results.  
    \code{tiskitpy} is just one of several modules to clean data and it is not needed to process the data: the point of this example is to
    demonstrate one approach to reading and cleaning the data and preparing the output files.
   After reading the data and metadata, the program plots the data, applies a simple rotation to the vertical channel to minimize its variance, then
   calculates transfer-function based cleaning functions  to generate the "cleaned for compliance" and "cleaned for earthquakes" data.
    It plots the original data and the cleaned waveforms  (Fig. \ref{fig:code_3_waveforms}), then the power spectral densities for all levels of cleaning (Fig: \ref{fig:code_3_PSDs}) and writes the "clean" data to a miniSEED file and the compliance calculation to a CSV file.


   Below is the section of the code for reading the input files:
   \begin{lstlisting}[caption=Reading the input files, label=code:readinput, language=Python]
from obspy import read, read_inventory, Stream
stream = read('../TRAINING/MAIN/MAIN_noisy.mseed', 'MSEED')
inv = read_inventory('../TRAINING/MAIN/MAIN_inv.xml', 'STATIONXML')
   \end{lstlisting}
   The code generates "cleaned for earthquakes" data (\code{stream\_rot\_12G}), "cleaned for compliance" data (\code{stream\_rot\_12}) and compliance (\code{ncompl\_calcd}).  Below is the section of the code for writing these out:
   \begin{lstlisting}[caption=Writing the output files, label=code:writeoutput, language=Python]
stream_rot_12G.write('MAIN_clean_for_eq.mseed', 'MSEED')
stream_rot_12.write('MAIN_clean_for_compl.mseed', 'MSEED')
ncompl_calcd.write('MAIN_compliance')
   \end{lstlisting}

   \code{ncompl\_calcd} is a \code{tiskitpy} object with its own ``write()`` method.  Other example codes show how to write the compliance output file without using \code{tiskitpy}.

    \subsubsection{Evaluating the results}
    \label{evaluation_example}
    The codes for evaluating the outputs are available in the \code{CODES/evaluation} directory.  You can either run each
    \code{eval\_*.py} file on the individual output files, or you can combine the files into a pseudo-submission directory and run \code{evaluate\_submission.py} on it.  For this example, collecting
    the outputs into one directory, adding the source code and running \code{evaluate\_submission.py} gives:

   \begin{lstlisting}[caption=Evaluating output files, label=evaluate, language=bash]
BRUIT-FM Challenge -- full evaluation of: ../examples/MAIN_out

  [FAIL] eval_manifest.py             score: 0.0
  [OK  ] eval_submission_files.py     score: 100.0
  [OK  ] eval_clean_for_eq.py         score: 97.5
  [OK  ] eval_clean_for_compl.py      score: 69.5
  [OK  ] eval_compliance.py           score: 30.3
  [OK  ] eval_code.py                 score: 93.3
   \end{lstlisting}

   \code{eval\_manifest} fails because we didn't create a manifest file for this test.
   
  \subsection{An example of cleaning the "real" challenge dataset}

\begin{figure*}[ht!]
		\centering
		\includegraphics[width=8.6cm]{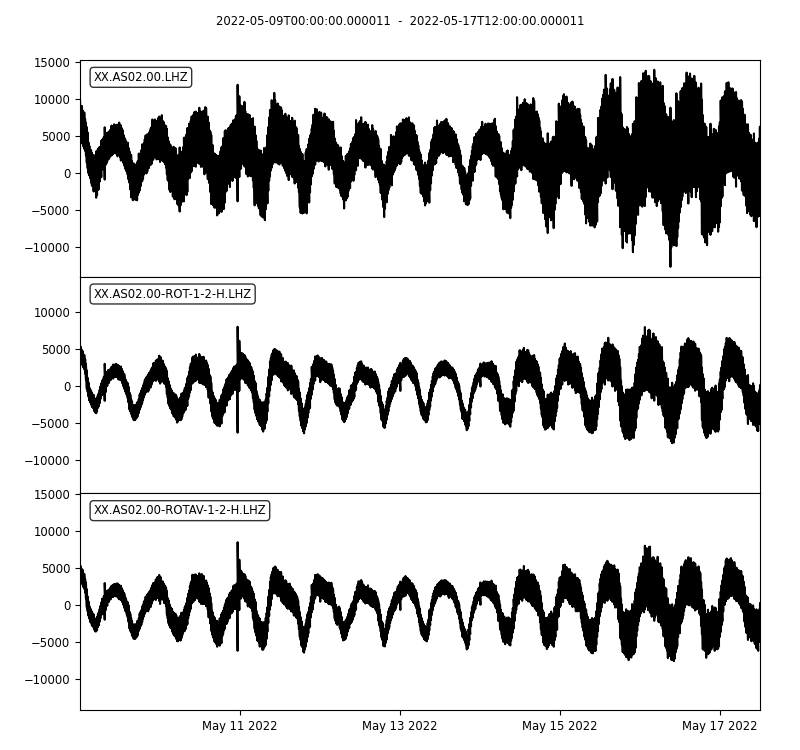} 
		\caption{Waveforms.  Top: Original data.  Middle : Cleaned with automatic \code{tiskitpy} data selection.  Bottom: Cleaned with manual data selection.}
		\label{fig:code_4_waveforms}
	\end{figure*}

\begin{figure*}[ht!]
		\centering
		\includegraphics[width=\textwidth]{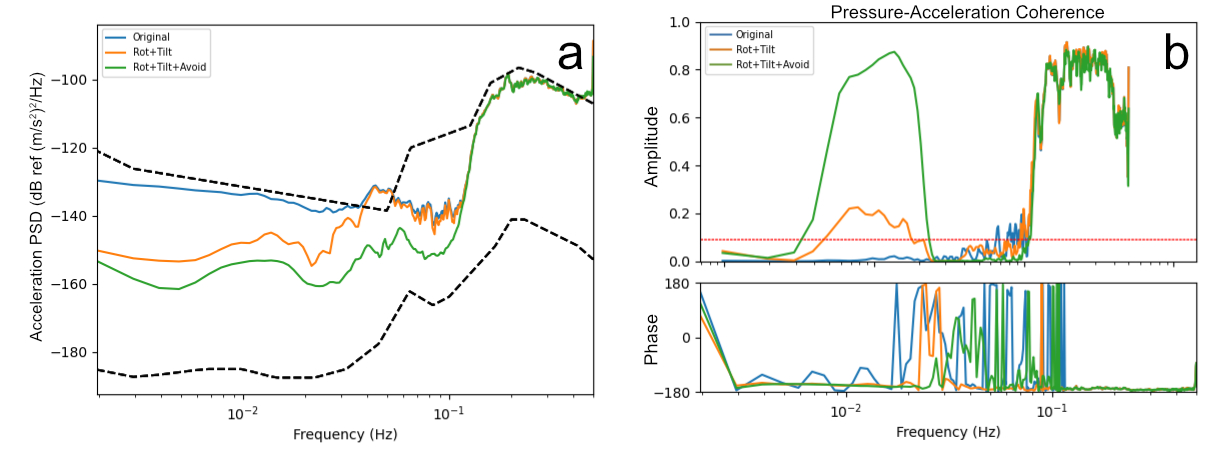} 
		\caption{(a) Vertical channel acceleration power spectral densities.  (b). Vertical-pressure coherences. Blue: original data.  Orange: Cleaned with automatic \code{tiskitpy} data selection.  Green: Cleaned with manual data selection.}
		\label{fig:code_4_PSDs_coherence}
	\end{figure*}

    Below are the results of the example code \code{CODE/step4\_clean\_avoid\_tiskitpy.py}, which reads in the \code{REAL/} challenge dataset and cleans it using \code{tiskitpy}, both automatically and with manual selection of which windows to use for the noise calculations.  
    Fig. \ref{fig:code_4_waveforms} shows waveforms for the original and cleaned data, while Fig: \ref{fig:code_4_PSDs_coherence} shows PSDs and coherences for these waveforms.  There is little visible difference between the automatic and manual window selection in the waveform plots, but the pressure-vertical coherence is significantly better for the manual window selection, allowing a more precise compliance calculation.

  \section{Submitting the Results}
  \label{section:submitting}

The zipped submission file should be submitted to the Challenge's \href{https://form.jotform.com/262153865166058}{open-source repository}, and a
 CSV file sampling the cleaned\_eq and cleaned\_compl outputs may be submitted to the Challenge's \href{https://kaggle.com/competitions/the-bruit-fm-open-data-noise-reduction-challenge}{Kaggle.com site}.  You can verify your submission directory and create the submission file using the script \code{CODES/submission/zip\_submission.py}.  Details on the submission files are specified in section \ref{submissions}.  The codes remain the property of each participant unless the submitted manifest file specifies otherwise.


		\begin{acknowledgements}
			This project is funded by the French national research agency (ANR) BRUIT-FM project (ANR-21-CE01-0031).  We thank Guilhem Barruol and Eleonore Stutzmann for project coordination and useful discussions.  We also thank Xongbo Zu for his input on microseism modeling, which should be included in a future version of the \code{SeafloorSynthetic} class \citep{crawford_2025_tiskitpy}.
		\end{acknowledgements}
		
		\section*{Data and code availability}
		The data are available at \citep{crawford_2025_bruitfmdata}.
        The \code{tiskitpy} package is available at \citep{crawford_2025_tiskitpy} or can be installed using the Python Package Index (command line: \code{pip install tiskitpy}).
		
		\section*{Competing interests}
		The authors have no competing interests.
		
		\section{Supplementary Information}

        \subsection{Directory structure}

        The files are organized as follows:
        \begin{itemize}
            \item \code{CHALLENGE/}: contains the challenge datasets,
                  one data file (\code{\{NAME\}\_noisy.mseed}) and one metadata file (\code{\{NAME\}\_inv.xml}) per subdirectory.  Subdirectories are:
                \begin{itemize}
                    \item \code{REAL/}: a real seafloor dataset (\code{\{NAME\}} = \code{REAL}).
                    \item \code{SYNTH/}: a synthetic seafloor dataset (\code{\{NAME\}} = \code{SYNTH}).
                \end{itemize}
              \item \code{TRAINING/}:  Training datasets.
                   Each dataset includes
                   \begin{itemize}
                       \item a data file (\code{\{NAME\}\_noisy.mseed}),
                       \item a metadata file (\code{\{NAME\}\_inv.xml}),
                       \item reference answer files. 
                       \begin{itemize}
                            \item \code{\{NAME\}\_REF\_clean\_eq.mseed},
                            \item \code{\{NAME\}\_REF\_clean\_compl.mseed},
                            \item \code{\{NAME\}\_REF\_compliance\_COUNTS.csv} and
                            \item \code{\{NAME\}\_REF\_compliance\_Pa-1.csv}
                       \end{itemize}
                   \end{itemize}
                   Subdirectories are:
             \begin{itemize}
                \item \code{MAIN/}: One 10-day training dataset (\code{\{NAME\}} = \code{MAIN})
                \item \code{MASS/}: One hundred 1-day training datasets (\code{\{NAME\}} = \code{MASS} for the inventory file,  \code{\{NAME\}} = \code{MASSxxx} for the data and answer files,
                      where \code{xxx} is a number from \code{000} to \code{099}).
              \end{itemize}
          \item \code{CODES/}: Useful Python codes.  Subdirectories are:
                \begin{itemize}
                    \item \code{examples/}: example codes for reading the data and writing the output files
                    \item \code{evaluation/}: The codes used to evaluate the submissions, with instructions on how to use them on training data outputs.
                    \item \code{submission/}: Codes to package the challenge outputs into the full submission file (ZIP file containing all answer files plus a manifest) and a "Kaggle" submission file (CSV file containing data sampled from the \code{clean\_eq} and/or \code{clean\_compl} output files).
                \end{itemize}
            \item \code{SUBMISSION_EXAMPLE/SUBMIT}: Contains an example submission dataset.  The data were generated from the MAIN example using the
            example codes \code{step3\_read\_write\_clean\_tiskitpy.py}
            for the \code{SYNTHETIC/} data and \code{step4\_clean\_avoid\_tiskitpy.py} for the 
                  \code{REAL/} data.  A submission ZIP file and Kaggle CSV file can be built using  \code{CODES/submission/zip\_submission.py}, which
                  also verifies that all necessary files are present.
       \end{itemize}

    \subsection{Input files}
    \label{fileformats}
     
      Input data files are in miniSEED format and are named \code{\{NAME\}\_noisy.mseed}.
      Inventory files are in StationXML format and are named \code{\{NAME\}\_inv.xml}. 

    \subsection{Answer files}
      Output data files should be in miniSEED format, use the same channel names as the input files, and be named
      \code{\{NAME\}\_clean\_for\_eq.mseed} if cleaned of everything except earthquakes  and \code{\{NAME\}\_clean\_for\_compl.mseed} if cleaned for compliance.
                
     Compliance values should be stored in a CSV file named
     \code{\{NAME\}\_compliance_\{UNITS\}.csv}, where \code{UNITS} is \code{Pa-1} or \code{COUNTS}.
     The file should have a first line with the fixed headers shown in Listing \ref{compl_out_files}, followed by as many lines as there are frequencies at which compliance was calculated. Frequencies are in Hz, compliance and uncertainty are in the appropriate units for the filename (COUNTS/COUNTS or Pa$^{-1}$) and phase is in degrees.  The `coherence` column may be omitted or left empty.
     Lines starting with \code{\#} are ignored.
    
   \begin{lstlisting}[caption=An example compliance output file, label=compl_out_files] 
# A comment line
frequencies;compliance;uncertainty;phase;coherence
0.001;1.0633e-05;1.33e-6;-30.341;0.854
0.002;4.4914e-05;1.42e-6;-64.724;0.890
0.003;8.5071e-05;1.21e-6;-96.279;0.911
   \end{lstlisting}

    \subsection{Submissions}
    \label{submissions}
    
       Answers should be collected into a directory containing at least two data files (one each for the \code{REAL} and \code{SYNTH} challenge data), a JSON-formatted manifest file named \code{MANIFEST.json}, and codes in a subdirectory named \code{code/}.
    Codes should be text files with appropriate suffixes (e.g., .txt, .md, .py, .cpp, ...).  If there is more than one code file, please include a \code{README.txt} or \code{README.md} file explaining each file.
The manifest file (example in Listing \ref{manifest_file}) should contain the fields: 
\begin{itemize}
    \item \code{"submission_format_version"} and \code{"dataset_version"}: leave as provided in \code{SUBMISSION\_EXAMPLE/SUBMIT/MANIFEST.json}.
    \item \code{nickname}: the name to display on the leaderboard.
    \item \code{licence}: the software licence for the code you provided (\code{"GPLv3"} is the most common free software licence, \code{null} means that you do not authorize any publication or reuse of the code).
    \item \code{"software_entry_point"} is the code file that calls or refers to any other files (if more than one). 
    \item \code{"notes"} are optional.
\end{itemize}

   \begin{lstlisting}[caption=Example manifest file, label=manifest_file] 
{
  "submission_format_version": "1.0",
  "dataset_version": "3.1",
  "nickname": "SilverSurfer",
  "licence": "GPLv3"
  "software_entry_point": "code/step3_read_write_clean_tiskitpy.py",
  "notes": "Optional participant comments"
}
    \end{lstlisting}

     \subsection{Example Codes}

     The \code{CODE/examples} directory contains example data processing codes.  The (\code{step*.py}) files show examples of reading, writing and processing the data (each step increasing in complexity). All but \code{step4} read from the "MAIN" training data.
     The \code{run_*.py} files show examples of applying a data cleaning routine to all of the \code{CHALLENGE} or \code{TRAINING} data.
     \begin{description}
        \item[\code{step0\_read\_write.py}] Read in miniseed data. Write out miniseed data and counts-based compliance.
        \item[\code{step1\_read\_write\_inv.py}] Read in miniseed data and StationXML inventory.  Write out miniseed data and "normalized" compliance.
        \item[\code{step2\_read\_write\_tiskitpy.py}] Read in miniseed data and StationXML inventory.  Write out miniseed data and "normalized" compliance, using \code{tiskitpy} \citep{crawford_2025_tiskitpy} to plot power spectral densities and coherences, and to write out the compliance.
        \item[\code{step3\_read\_write\_clean\_tiskitpy.py}] Read in miniseed data and StationXML inventory.  Write out miniseed data and normalized compliance, using \code{tiskitpy} to automatically clean the data.
        \item[\code{step4\_clean\_avoid\_tiskitpy.py}] Read in miniseed data and StationXML inventory from the "REAL" challenge dataset.  Write out miniseed data and normalized compliance , using \code{tiskitpy} to clean the data, with manually-selected timespans.
        \item[\code{run\_training.py}] Runs a cleaning function on all of the training files, writing the outputs to \code{OUTPUTS/TRAINING}.
        \item[\code{run\_challenge.py}] Runs a cleaning function  on all of the challenge files, writing the outputs to \code{OUTPUTS/CHALLENGE}.
    \end{description}

		\bibliography{mybibfile}

	\end{document}